\documentclass[12pt,draftclsnofoot,onecolumn]{IEEEtran}

\usepackage{cite}      

\usepackage{graphicx}  

\usepackage{psfrag}    
\usepackage{amssymb}
\newcommand{\real}{{\mathbb{R}}}

\def\ns{\hspace{-1mm}}

\def\gI{{\cal I}}

\def\gP{{\cal P}}

\def\gV{{\cal V}}

\begin{document}
%
\title{\Large 
A Numerical Example about the Geometric Approach to the Output Regulation 
Problem with Stability for Linear Switching Systems}
%
%
%
\author{Elena~Zattoni$^\ast$ 
        Anna Maria Perdon$^\diamond$ 
        Giuseppe Conte$^\diamond$
\thanks{$^\ast$E.~Zattoni is with the Department of Electrical, Electronic and Information
Engineering ``Guglielmo Marconi'', Alma Mater Studiorum $\cdot$ University of Bologna, 
Viale Risorgimento~2, 40136 Bologna, Italy. E-mail: elena.zattoni@unibo.it}
\thanks{$^\diamond$A.~M.~Perdon and G.~Conte are with the Department of Information Engineering, 
Polytechnic University of Marche, Via~Brecce Bianche, 60131 Ancona, Italy. 
E-mail: perdon@univpm.it, gconte@univpm.it}}

\maketitle
\begin{abstract}
\noindent
This note presents a numerical example worked out in order to illustrate the 
solution to the output regulation problem with quadratic stability for linear 
switching systems derived in \cite{Zattoni-PC-2013}. 
\end{abstract}


%
\IEEEpeerreviewmaketitle

\thispagestyle{empty}
%
%
\section{Introduction}
In \cite{Zattoni-PC-2013}, the classical output regulation problem for linear 
time-invariant systems (see, e.g., \cite{Francis-W-1976, Francis-1977}) has 
been given an extended formulation in the case of linear switching systems. The 
solution of the problem with stability has been achieved through the geometric 
approach \cite{Wonham-1985, Basile-M-1992}, along the lines first proposed in 
\cite{Marro-1996} for linear time-invariant systems. The main feature of the 
geometric solution developed in \cite{Zattoni-PC-2013} is allowing the structural 
issue (i.e., the requirement that the regulation error goes to zero uniformly under 
arbitrary switching) to be dealt with separately from the stability issue (i.e., 
the requirement that the regulation loop satisfies specific stability conditions 
under arbitrary switching). In this way, different stability specifications, such as 
asymptotic stability, exponential stability, or quadratic stability, only impact on few, 
precisely-defined, aspects of the solution. In particular, the synthesis procedure 
presented in \cite{Zattoni-PC-2013} refers to quadratic stability and makes an extensive 
use of linear matrix inequalities \cite{Boyd-EGFB-1994, Feron-1996, Garcia-BA-1996, 
ElGhaoui-N-2000}. The scope of this technical note is to help the reader in the 
implementation of the proposed synthesis procedure by illustrating the single steps 
with a numerical example.
%
%
\section{Notation}
The symbol $\real$ stands for the set of real numbers. Matrices and linear maps are 
denoted by upper-case letters, like $A$. The image of $A$ is denoted by ${\rm Im}\,A$. 
The transpose of $A$ is denoted by $A^\top$. Vector spaces and subspaces are denoted 
by calligraphic letters, like~$\gV$. The symbols $I_n$, $O_{m\,{\times}\,n}$, and $0_n$ 
are respectively used for the identity matrix of dimension $n$, the $m\,{\times}\,n$ 
zero matrix, and the $n$-dimensional zero vector (subscripts are omitted when the 
dimensions can be inferred from the context). The symbol $M\,{>}\,0$, where 
$M\,{\in}\,\real^{n\times n}$ is symmetric, means that $M$ is positive-definite:
i.e., $x^\top M\,x\,{>}\,0$ for all nonzero $x\,{\in}\,\real^n$. Similarly, $M\,{<}\,0$
means that $M$ is negative-definite. The symbol $\lambda_{\rm max}(M)$ denotes the maximal
eigenvalue of the matrix $M\,{=}\,M^\top$. Similarly, $\lambda_{\rm min}(M)$ denotes the
minimal eigenvalue of $M$. 
%
%
\section{A Numerical Example}
\label{sect_introduction}
%
%
%
%
%
%
The aim of this section is to illustrate a computational framework for the 
synthesis procedure developed in \cite{Zattoni-PC-2013}, with the help of 
a numerical example. The basic tools that will be used are the subspace 
computation algorithms of the Geometric Approach Toolbox, first appeared 
in \cite{Basile-M-1992} and now available on-line in an upgraded version, 
and the LMI solvers of the Robust Control Toolbox \cite{Balas-CPS-2012}. 
The variables will be displayed in scaled fixed point format with five digits, 
although the computation will be made in floating point precision. 
%
\par
Let $\Sigma_{\sigma(t)}$, defined by (1) in \cite{Zattoni-PC-2013}, be a 
discrete-time sample-data linear switching system, with sampling time 
$T_s\,{=}\,0.1$\,s. Let $\gI\,{=}\,\{1,2\}$ and
\begin{eqnarray*}
A_1\ns&\ns=\ns&\ns 
\left[\begin{array}{cccccc}
       0.6 &       0 &       0 &       0 &  0.1    &       0\\
    	 1 &  	   1 &    -0.3 &       0 & -0.2    &       0\\
         0 &  	 0.4 &       0 &       0 &  1      &       0\\
         0 &       0 &       0 &     0.9 &  0.25   &    -0.4\\
         0 &       0 &       0 &       0 &  0.3    &       0\\
         0 &       0 &       0 &       0 & -0.2    &     0.7
\end{array}\right],\quad
B_1 =
\left[\begin{array}{ccc}
        -2  &      1 &       0\\
        -1  &      0 &       0\\
         0  &      0 &       1\\
         0  &      1 &       0\\
         3  &      0 &       0\\
        0.4 &      1 &       1
\end{array}\right],\\
C_1\ns&\ns=\ns&\ns 
\left[\begin{array}{cccccc}
         0 &       0 &     2.8 &       0 &  0    &       0\\
    	 0 &  	   0 &       0 &       1 &  0    &       0
\end{array}\right],\\
A_2\ns&\ns=\ns&\ns 
\left[\begin{array}{cccccc}
       0.6 &       0 &       0 &       0 & -0.3    &       0\\
    	 1 &  	   1 &    -0.3 &       0 &  0      &       0\\
         0 &  	 0.4 &       0 &       0 &  0      &       0\\
         0 &       0 &       0 &     0.9 &  0      &    -0.4\\
         0 &       0 &       0 &       0 &  0.5    &       0\\
         0 &       0 &       0 &       0 &  0      &     0.7
\end{array}\right],\quad
B_2 =
\left[\begin{array}{ccc}
         1  &      1 &       0\\
         0  &      0 &       0\\
         0  &      0 &       1\\
         0  &      1 &       0\\
         1  &      0 &       0\\
         0  &      1 &       1
\end{array}\right],\\
C_2\ns&\ns=\ns&\ns 
\left[\begin{array}{cccccc}
         0 &       0 &     2.8 &       0 &  0    &       0\\
    	 0 &  	   0 &       0 &       1 &  0    &       0
\end{array}\right].
\end{eqnarray*}
Let the exogenous system $\Sigma_{g,\sigma(t)}$, defined by (3) in \cite{Zattoni-PC-2013}, 
have the following matrices
\begin{eqnarray*}
A_{g,1}\ns&\ns=\ns&\ns A_{g,2} =
\left[\begin{array}{cccc}
	1 & 1 & 0 & 0\\ 
	0 & 1 & 0 & 0\\
	0 & 0 & 1 & 1\\
	0 & 0 & 0 & 1
\end{array}\right],\\
E_{g,1}\ns&\ns=\ns&\ns E_{g,2} =
\left[\begin{array}{cccc}
	1 & 0 & 0 & 0\\ 
	0 & 0 & 1 & 0\\
\end{array}\right].
\end{eqnarray*}
Namely, the internal model of the ramp signal is replicated in the exogenous 
system dynamics a number of times equal to the number of the outputs of the 
to-be-controlled system, so that independent reference signals can be obtained 
for each output, based on the exosystem initial state. 
\par
Assumption~1 in \cite{Zattoni-PC-2013} is satisfied, since the switching system 
$\Sigma_{\sigma(t)}$ is quadratically stable under arbitrary switching. In fact, 
the LMIs $A_i^\top\,Q\,A_i\,{-}\,Q\,{<}\,0$ hold for all $i\,{\in}\,\gI$ with 
$Q$ given, e.g., by
\[
Q= 
\left[\begin{array}{cccccc}
    6.5135 &  0.2046 & -0.2647 & -0.0873 & -0.7930 & -0.0446\\
    0.2046 &  1.0830 & -1.0809 &  0.0752 &  0.0876 & -0.1392\\
   -0.2647 & -1.0809 &  2.8705 & -0.1276 & -0.2097 &  0.2743\\
   -0.0873 &  0.0752 & -0.1276 &  2.0635 &  0.4823 & -0.6029\\
   -0.7930 &  0.0876 & -0.2097 &  0.4823 &  6.5706 & -0.7474\\
   -0.0446 & -0.1392 &  0.2743 & -0.6029 & -0.7474 &  6.1649
\end{array}\right],
\]
which is symmetric and positive-definite. Moreover, Assumption~2 in \cite{Zattoni-PC-2013} 
is satisfied by the switching system $\Sigma_{e,\sigma(t)}$, defined according to (5)--(7)
in \cite{Zattoni-PC-2013}. In fact, $\Sigma_{e,\sigma(t)}$ is quadratically stabilizable
under arbitrary switching by the linear output injection matrices
\[
G_{e,1}=\left[\begin{array}{cc}
	    0.0116 & -0.0035\\
	    0.0490 & -0.0183\\
	    0.0469 &  0.0082\\
	   -0.0057 &  0.0525\\
	    0.0043 &  0.0036\\
	   -0.0043 & -0.0248\\
	   -0.6393 &  0.0644\\
	   -0.1291 &  0.0184\\
	   -0.0118 & -1.3235\\
	   -0.0008 & -0.3628
	\end{array}\right],\quad
G_{e,2}=\left[\begin{array}{cc}
	    0.0074 & -0.0078\\
	    0.0500 & -0.0157\\
	    0.0346 & -0.0016\\
	   -0.0084 &  0.0500\\
	    0.0042 &  0.0059\\
	   -0.0016 & -0.0229\\
	   -0.6393 &  0.0649\\
	   -0.1293 &  0.0185\\
	   -0.0113 & -1.3235\\
	   -0.0007 & -0.3627
	\end{array}\right],
\]
which have been derived through the solution of the LMIs (A.8),
as specified in Appendix~A of \cite{Zattoni-PC-2013}. Hence, in 
order to solve Problem~1 in \cite{Zattoni-PC-2013} (i.e., the 
autonomous regulator problem with quadratic stability or, briefly,
ARPQS), Conditions~(i), (ii) of Theorem~2 in \cite{Zattoni-PC-2013} 
are checked. The maximal robust controlled invariant subspace 
$\gV^\ast_R$ is given by 
\[
\gV^\ast_R={\rm Im}\,
	\left[\begin{array}{cccccccc}
		 1\ & 0\ & 0\	   & 0\		& 0\ &\ 0\ & 0\ & 0\\ 
		 0\ & 1\ & 0\ 	   & 0\ 	& 0\ &\ 0\ & 0\ & 0\\
		 0\ & 0\ &-0.3363\ & 0\		& 0\ &\ 0\ & 0\ & 0\\
		 0\ & 0\ & 0\      & 0.7071\	& 0\ &\ 0\ & 0\ & 0\\
		 0\ & 0\ & 0\      & 0\ 	& 1\ &\ 0\ & 0\ & 0\\
		 0\ & 0\ & 0\      & 0\ 	& 0\ &\ 1\ & 0\ & 0\\
		 0\ & 0\ &-0.9417\ & 0\		& 0\ &\ 0\ & 0\ & 0\\
		 0\ & 0\ & 0\      & 0\		& 0\ &\ 0\ &-1\ & 0\\
		 0\ & 0\ & 0\      & 0.7071\	& 0\ &\ 0\ & 0\ & 0\\
		 0\ & 0\ & 0\	   & 0\		& 0\ &\ 0\ & 0\ & 1
	\end{array}\right].
\]
The subspace $\gP$ is defined as 
\[
\gP\,{=}\,{\rm Im}\,
	\left[\begin{array}{c} 
		I_6\\ 
             	O_{4\times 6}
	\end{array}\right],
\]
according to (19) in \cite{Zattoni-PC-2013}. Therefore, Condition~(i) 
of Theorem~2 in \cite{Zattoni-PC-2013} is satisfied. Moreover, taking 
a suitable set of state feedbacks satisfying (39) of Lemma~2 in 
\cite{Zattoni-PC-2013}, one gets that also Condition~(ii) of 
Theorem~2 in \cite{Zattoni-PC-2013} is met. In particular, a subspace 
$\gV$, computed along the lines sketched in the proof of Theorem~2 in
\cite{Zattoni-PC-2013}, and, therefore, satisfying Conditions~(i), (ii) 
of Theorem~1 in \cite{Zattoni-PC-2013} is 
\[
\gV={\rm Im}\,
	\left[\begin{array}{cccc}
    		0.1009 &  0	 & -0.9196 & -0.1875\\
		0.8661 & -0.1326 & -0.9219 & -2.5469\\
		0.3363 &  0	 &  0	   &  0	\\
		0      & -0.7071 &  0	   &  0	\\
		0      &  0	 &  0	   &  0	\\
		0.1009 &  0.1768 & -1.1875 &  2.3125\\
		0.9417 &  0	 &       0 &       0\\
		0      &  0   	 & -1	   &       0\\
		0      & -0.7071 &       0 &       0\\
		0      &  0      &       0 & -1
	\end{array}\right].
\]
Consequently, a set $\{F_{e,i},\,i\,{\in}\,\gI\}$ of state feedbacks, defined 
according to Lemma~1 in \cite{Zattoni-PC-2013}, consists of the matrices 
{\small
\begin{eqnarray*}
F_{e,1} = \\
& &\hspace{-18mm}
\left[\begin{array}{cccccccccc}
 -0.0998  & 0.0551 & -0.1138 & -0.0053 &  0.3356 &  0.0230 & -0.0018 &  0.0137 &  0.0007 & -0.0685\\
 -0.2954  & 0.2055 & -0.4266 & -0.0302 &  0.9029 &  0.1199 &  0.0251 &  0.4149 &  0.0216 & -0.1157\\
  0.1318  &-0.1232 &  0.2606 & -0.0053 & -0.5643 &  0.0546 & -0.0104 & -0.0841 & -0.0330 & -0.6033
\end{array}\right]\!,
\end{eqnarray*}
\begin{eqnarray*}
F_{e,2} = \\
& &\hspace{-18mm}
\left[\begin{array}{cccccccccc}
 0.1502 & -0.1593 &  0.3113 &  0.0374 &  0.3058 & -0.1429 &  0.0346 &  0.1784 & -0.0433 &  0.0470\\
-0.0537 &  0.1395 & -0.2776 & -0.0426 & -0.0045 &  0.1620 &  0.0021 &  0.2035 &  0.0569 &  0.1044\\
 0.3953 & -0.2748 &  0.5710 &  0.0130 &  0.0096 & -0.0227 & -0.0019 & -0.0949 & -0.0421 & -0.4456
\end{array}\right]\!.\\
\end{eqnarray*}
}
Instead, a set $\{G_{e,i},\,i\,{\in}\,\gI\}$ of output injections, defined as 
in the if-part of the proof of Theorem~1 in \cite{Zattoni-PC-2013}, consists 
of the matrices $G_{e,1}$ and $G_{e,2}$ previously computed. Finally, the
matrices of the switching regulator $\Sigma_{r,\sigma(t)}$, defined by 
(8) in \cite{Zattoni-PC-2013}, are determined according to (21)--(22) in 
\cite{Zattoni-PC-2013}.
\par
\begin{figure}[t]
\begin{center}
\includegraphics[width=\columnwidth]{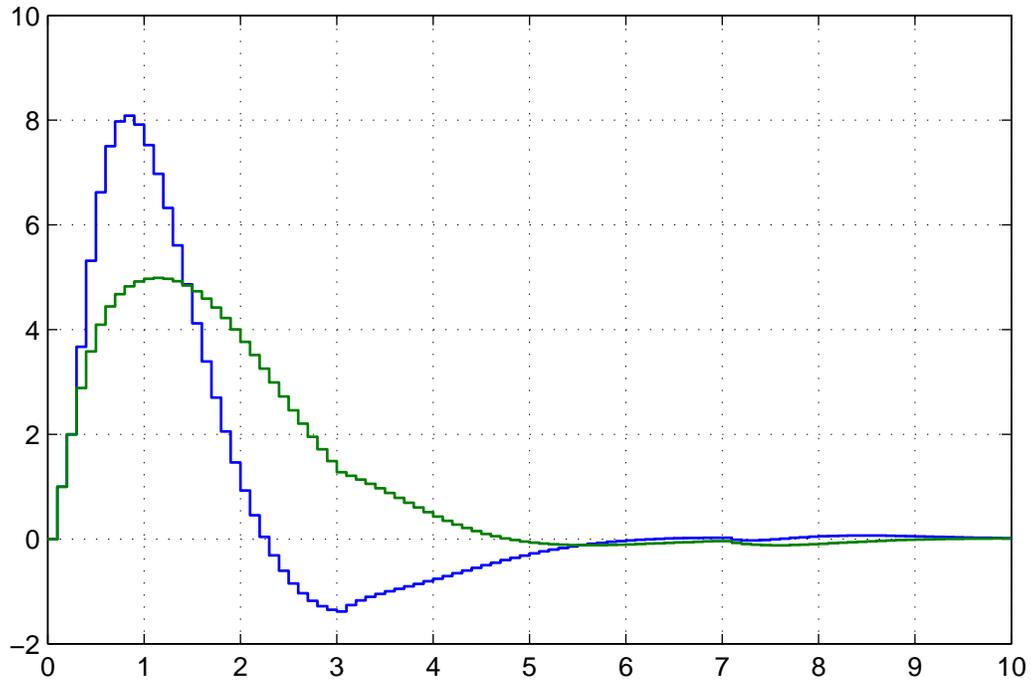}
\caption{Graphs of the error -- amplitude vs.\ time (s)}
\label{error}
\end{center}
\end{figure}

A simulation on the given switching system $\Sigma_{\sigma(t)}$ and the given exogenous 
system $\Sigma_{g,\sigma(t)}$ is run with the following data. The time goes from $0$\,s to 
$10$\,s, so that, the total number of samples is $100$. The switching signal $\sigma(t)$ 
is defined by
\[
\sigma(t)=\left\{\begin{array}{ll}
		1,\quad & \mbox{with}\ \ t=0,\ldots,29,\\
		2,\quad & \mbox{with}\ \ t=30,\ldots,69,\\
		1,\quad & \mbox{with}\ \ t=70,\ldots,99.
	 \end{array}\right.
\]
Namely, the active mode of $\Sigma_{\sigma(t)}$ is $\Sigma_1$, between $0$\,s 
and $2.9$\,s as well as between $7$\,s and $9.9$\,s, while the active mode 
is $\Sigma_2$ between $3$\,s and $6.9$\,s. Figure~\ref{error} shows the graphs 
of the error for the two outputs. The graphs clearly reveal the different dynamics
before and after the time $3$\,s, when the first switch occurs, while the effect 
of the second switch, at the time $7$\,s, is less evident, mainly because of the 
smaller values of the error, which goes to zero as the time increases.
%
%
%
%
%
%
\section{Conclusion}
The implementation of the procedure for synthesizing a switching compensator achieving
asymptotic tracking of the reference signal and quadratic stability of the closed-loop
presented in \cite{Zattoni-PC-2013} has been illustrated by means of a numerical example.
References have been provided for the computational tools that have been used.
\bibliographystyle{IEEEtran}
\bibliography{aut12-0550arXiv}

\begin{thebibliography}{10}
\providecommand{\url}[1]{#1}
\csname url@samestyle\endcsname
\providecommand{\newblock}{\relax}
\providecommand{\bibinfo}[2]{#2}
\providecommand{\BIBentrySTDinterwordspacing}{\spaceskip=0pt\relax}
\providecommand{\BIBentryALTinterwordstretchfactor}{4}
\providecommand{\BIBentryALTinterwordspacing}{\spaceskip=\fontdimen2\font plus
\BIBentryALTinterwordstretchfactor\fontdimen3\font minus
  \fontdimen4\font\relax}
\providecommand{\BIBforeignlanguage}[2]{{%
\expandafter\ifx\csname l@#1\endcsname\relax
\typeout{** WARNING: IEEEtran.bst: No hyphenation pattern has been}%
\typeout{** loaded for the language `#1'. Using the pattern for}%
\typeout{** the default language instead.}%
\else
\language=\csname l@#1\endcsname
\fi
#2}}
\providecommand{\BIBdecl}{\relax}
\BIBdecl

\bibitem{Zattoni-PC-2013}
E.~Zattoni, A.~M. Perdon, and G.~Conte, ``The output regulation problem with
  stability for linear switching systems: {A} geometric approach,''
  \emph{Automatica}, vol.~49, no.~10, pp. 2953--2962, October 2013,
  http://dx.doi.org/10.1016/j.automatica.2013.07.005.

\bibitem{Francis-W-1976}
B.~A. Francis and W.~M. Wonham, ``The internal model principle of control
  theory,'' \emph{Automatica}, vol.~12, pp. 457--463, 1976.

\bibitem{Francis-1977}
B.~A. Francis, ``The linear multivariable regulator problem,'' \emph{SIAM
  Journal on Control and Optimization}, vol.~15, no.~3, pp. 486--505, May 1977.

\bibitem{Wonham-1985}
W.~M. Wonham, \emph{Linear Multivariable Control: A Geometric Approach},
  3rd~ed.\hskip 1em plus 0.5em minus 0.4em\relax New York: Springer-Verlag,
  1985.

\bibitem{Basile-M-1992}
G.~Basile and G.~Marro, \emph{Controlled and Conditioned Invariants in Linear
  System Theory}.\hskip 1em plus 0.5em minus 0.4em\relax Englewood Cliffs, New
  Jersey: Prentice Hall, 1992.

\bibitem{Marro-1996}
G.~Marro, ``Multivariable regulation in geometric terms: Old and new results,''
  in \emph{Colloquium on Automatic Control}, ser. Lecture Notes in Control and
  Information Sciences, C.~Bonivento, G.~Marro, and R.~Zanasi, Eds.\hskip 1em
  plus 0.5em minus 0.4em\relax Berlin / Heidelberg: Springer, 1996, vol. 215,
  pp. 77--138.

\bibitem{Boyd-EGFB-1994}
S.~Boyd, L.~El~Ghaoui, E.~Feron, and V.~Balakrishnan, \emph{Linear Matrix
  Inequalities in System and Control Theory}, ser. {SIAM} Studies in Applied
  and Numerical Mathematics.\hskip 1em plus 0.5em minus 0.4em\relax
  Philadelphia, PA: Society for Industrial and Applied Mathematics, 1994.

\bibitem{Feron-1996}
E.~Feron, ``Quadratic stabilizability of switched linear systems via state and
  output feedback,'' Center for Intelligent Control Systems, Providence, RI,
  Tech. Rep. CICS-P-468, 1996.

\bibitem{Garcia-BA-1996}
G.~Garcia, J.~Bernussou, and D.~Arzelier, ``Stabilization of an uncertain
  linear dynamic system by state and output feedback: A quadratic
  stabilizability approach,'' \emph{International Journal of Control}, vol.~64,
  no.~5, pp. 839--858, 1996.

\bibitem{ElGhaoui-N-2000}
L.~El~Ghaoui and S.-I. Niculescu, Eds., \emph{Advances in Linear Matrix
  Inequality Methods in Control}, ser. Advances in Design and Control.\hskip
  1em plus 0.5em minus 0.4em\relax Philadelphia, PA: Society for Industrial and
  Applied Mathematics, 2000.

\bibitem{Balas-CPS-2012}
G.~Balas, R.~Chiang, A.~Packard, and M.~Safonov, ``{LMI Solvers},'' in
  \emph{Robust Control Toolbox\texttrademark -- Getting Started Guide
  {R2012b}}.\hskip 1em plus 0.5em minus 0.4em\relax Natick, {MA}: The MathWorks
  Inc., 2012, pp. \mbox{1--22}.

\end{thebibliography}
\end{document}